\documentclass[10pt]{iopart}
\usepackage{graphicx}
\usepackage{amssymb}
\usepackage{cite}

\begin{document}

\title{On the number and spin of photons in classical electromagnetic field}
\author{R M Feshchenko$^1$\footnote{Present address: P.N. Lebedev Physical Institute of RAS, 53 Leninski Prospect, Moscow, 119991,  Russia}, and A V Vinogradov$^1$}

\address{$^1$ P.N. Lebedev Physical Institute of RAS, 53 Leninski Prospect, Moscow, 119991,  Russia}

\ead{rusl@sci.lebedev.ru}

\begin{abstract}\noindent
A relativistically invariant expression for the number of photons in free classical electromagnetic field through the currents, that created the field, is derived based on the formula for the total energy--momentum of the field. It is demonstrated that it corresponds to the classical limit of the photon number operator known from the quantum electrodynamics. An expression for the total spin moment of the classical electromagnetic field is derived and it is shown that it can be interpreted in terms of spin moments of elementary photons. Similar to the total number of photons the classical spin moment is the limit of the quantum operator of the total spin. It is revealed that the total number of photons as well as the total spin moment are related to each other through a second order tensor, which can be expressed through currents that created the field.
\end{abstract}


\submitto{\EJP}
\maketitle
\ioptwocol

\section{Introduction}
In his 1966 work Zeldovich stated that in the free classical electromagnetic field there exists a conserved quantity, which has the meaning of the number of photons in this field \cite{Zeldovich1966}. However that expression as written in \cite{Zeldovich1966} was not explicitly relativistically invariant. In addition it was not clear how it was related to the currents that emitted the electromagnetic field. Those shortcomings led to some difficulties when calculation of the number of photons contained in a spatially finite field configuration (inside a Fabry--Perot cavity) was attempted in two different reference frames \cite{avron1999number,mcdonaldlorentz}. Thus a re-evaluation of the concept of photon number in the classical electromagnetic field was needed.

In the previous paper \cite{FeshchenkoVinogradov2018} we derived a relativistically invariant expression for the Zeldovich invariant and expressed it through the currents that radiated the field. We also considered a couple of toy models where the meaning of the photon number as applied to the classical electromagnetic radiative fields became clearer. 

In the present work we re-derive the expression for the total number of photons in a more systematic way and demonstrate how Zeldovich expression relates to the total photon number operator known from the quantum electrodynamics. We also investigate the relationship between the total number of photons and the total spin angular momentum of the free electromagnetic field and its quantum operator. The obtained expressions have a methodological value and may help those who study the classical and quantum electrodynamics in understanding the concept of photon and how it relates to the spin angular momentum of the electromagnetic field and to its energy--momentum as well as to their quantum counterparts.

\section{Total energy--momentum and spin angular momentum radiated by a finite system of currents}
Let us re-derive the expression for the total number of photons obtained in our previous work but in a different way \cite{FeshchenkoVinogradov2018}. The 4D Fourier transform of the potentials (in Lorentz gauge and assuming that speed of light $c=1$) of the radiative (free) electromagnetic field $A^l(k)$ emitted by 4D current $j^l(k)$, which are considered to occupy a finite spatial and temporal domain, can be written as in \cite{fedotov2007exact}
\begin{eqnarray}
A^i(k)&=\left(\Pi^{ret}(k)-\Pi^{adv}(k)\right)j^i(k)\nonumber\\
&=\left[\frac{4\pi}{(k^0-i\varepsilon)^2-{\mathbf{k}}^2}-\frac{4\pi}{(k^0+i\varepsilon)^2-{\mathbf{k}}^2}\right]j^i(k),\label{1a}
\end{eqnarray}
where $\epsilon\to0$, $k=(k^0, \mathbf{k})$ is a 4D wave vector, $k^0$ is the frequency, $\mathbf{k}$ is the 3D wave vector,
\begin{eqnarray}
A^i(k)=\int A^i(x)e^{ikx}\,d^4x,\label{1b1}\\
j^i(k)=\int j^i(x)e^{ikx}\,d^4x,\label{1b2}
\end{eqnarray}
where $x=(t=x^0, \mathbf{r})$ are 4D coordinates, $j^k=(\rho,\mathbf{j})$ is the 4D current ($\rho$ is the volume charge density and $\mathbf{j}$ is the 3D vector current) whereas $\Pi^{ret}$ and $\Pi^{adv}$ are the retarded and advanced propagators, respectively. Taking the limit $\epsilon\to0$ in (\ref{1a}) one can obtain that
\begin{equation}
\label{1c}
A^i(k)=-i8\pi^2\mathrm{sign}(k^0)\delta(k^2)j^i(k),
\end{equation}
where $k^2=k^\nu k_\nu$ (summation is implied for the repeated upper and lower indexes), $\mathrm{sign}(x)=2\theta(x)-1$ is the sign function and $\theta(x)$ is theta (step) function. Since in formula (\ref{1c}) the positive and negative frequency parts have the same absolute values it is sufficient to consider only the positive frequency part 
\begin{equation}
\label{1d}
A^i(k)=-i8\pi^2\theta(k^0)\delta(k^2)j^i(k).
\end{equation}
The contribution of the negative frequency part to the total energy--momentum and to the total angular momentum is equal to that of the positive frequency part. 

The total energy--momentum $P^i$ of the electromagnetic field can then be found by integrating the energy--momentum tensor $T^{ik}$ of the electromagnetic field over a hyperplane $\cal S$ orthogonal to some unity 4D vector $n$ ($n^2=1$) \cite{landau2013classical}
\begin{equation}
\label{1e}
P^i=\int\limits_{\cal S}{T}^{ik}\,dS_k,
\end{equation}
where \cite{landau2013classical}
\begin{equation}
\label{1f}
T^{ik}=-\frac{1}{4\pi}\left(F^{il}{F^k}_l-\frac{g^{ik}}{4}F^{lm}F_{lm}\right)
\end{equation}
and $g^{ik}=\mathrm{diag}(1,-1,-1,-1)$ is the Minkowski space metric tensor. Electromagnetic field tensor $F^{ik}$ is \cite{landau2013classical}
\begin{equation}
\label{1g}
F^{ik}=\frac{\partial A^k}{\partial x_i}-\frac{\partial A^i}{\partial x_k},
\end{equation}
where potentials $A^i(x)$ are defined here as the inverse Fourier transform of expression (\ref{1d}) containing only positive frequencies. Using (\ref{1d})--(\ref{1g}) and assuming that the hyperplane ${\cal S}$ lies outside the domain where the currents are present, the total energy--momentum radiated by a finite system of currents can be expressed through the positive frequency 4D Fourier harmonics of the 4D current as 
\begin{equation}
\label{1h}
P^i=-\frac{1}{2\pi^2}\int\theta(k^0)\delta(k^2)k^ij^l(k)j^*_l(k)\,d^4k. 
\end{equation}
The tensor of total radiated spin momentum $S^{im}$ can be obtained in the same way as the total energy--momentum vector. Using the definition of total angular momentum of the electromagnetic field \cite{landau2013classical}
\begin{equation}
\label{1h1}
M^{im}=\int\limits_{\cal S}\left[{x^i T}^{mk}-{x^m T}^{ik}\right]\,dS_k,
\end{equation}
and substituting (\ref{1d})--(\ref{1g}) in (\ref{1h1}) one obtains that
\begin{eqnarray}
M^{im}&=&-\frac{1}{\pi^2}\mathrm{Im}\int\theta(k^0)\delta(k^2)j^i(k){j^m}^*(k)\,d^4k\nonumber\\
&+&\frac{1}{2\pi^2}\mathrm{Im}\int\theta(k^0)\delta(k^2)k^m j^s(k)\frac{\partial{j_s}^*(k))}{\partial k_i}\,d^4k.\label{1h2}
\end{eqnarray}
It can be shown that in (\ref{1h2}) the first term corresponds to the spin part of the angular momentum and the second part corresponds to the orbital momentum. Therefore the total radiated spin momentum 
\begin{equation}
\label{1i}
S^{im}=-\frac{1}{\pi^2}\mathrm{Im}\int\theta(k^0)\delta(k^2)j^i(k){j^m}^*(k)\,d^4k 
\end{equation}
depends only on components of currents but not on their derivatives by $k^m$. The spin momentum defined by (\ref{1i})  is conserved in case of the electromagnetic field in free space \cite{cameron2017chirality}. One can also observe that expression (\ref{1i}) is in some way similar to (\ref{1h}). 

Now using the definition of angular momentum tensor \cite{landau2013classical} and integrating by $k^0$ in (\ref{1i}) one can obtain for the radiated 3D spin momentum vector
\begin{equation}
\label{1j}
\mathbf{S}=-\frac{1}{2\pi^2}\mathrm{Im}\int\frac{\mathbf{j}(|\mathbf{k}|,\mathbf{k})\times\mathbf{j}^*(|\mathbf{k}|,\mathbf{k})}{|\mathbf{k}|}\,d^3k
\end{equation}
which can be transformed into the standard expression \cite{van1994spin} for spin angular momentum by switching to the coordinate representation (see below) and using the 4D Gauss theorem.

\section{Zeldovich formula}
The total number of photons can be introduced into the classical electromagnetic theory if formula (\ref{1h}) is re-written as 
\begin{equation}
\label{2a}
P^i=\int{p^i}n(k)\,d^4k,
\end{equation}
where $p^i=\hbar k^i$ is the 4D momentum of a single photon and $n(k)$ is a quantity, which can be interpreted as the classical photon phase-space density. From (\ref{1h}) it follows that
\begin{equation}
\label{2b}
n(k)=-\frac{1}{2\pi^2\hbar}\theta(k^0)\delta(k^2)j^l(k)j^*_l(k),
\end{equation}
and therefore the total number of photons is
\begin{equation}
\label{2c}
N=-\frac{1}{2\pi^2\hbar}\int\theta(k^0)\delta(k^2)j^l(k)j^*_l(k)\,d^4k,
\end{equation} 
which can be also re-written in 3D form as
\begin{equation}
\label{2c1}
N=\frac{1}{4\pi^2\hbar}\int\frac{|\mathbf{j}_{\perp}(|\mathbf{k}|,\mathbf{k})|^2}{|\mathbf{k}|}\,d^3k,
\end{equation}
where it was taken into account that since any 4D current must satisfy the continuity equation $j^\nu k_\nu=\rho k^0-\mathbf{jk}=0$ the charge density and the longitudinal component of $\mathbf{j}$ can be both eliminated and the final expression depends only on two component vector $\mathbf{j}_\perp$ orthogonal to $\mathbf{k}$. Moreover taking into account that because expression (\ref{1d}) is invariant relative to the gauge transformation $\delta(k^2)j^i(k)\to\delta(k^2)j^i(k)+\delta(k^2)c(k)k^i$ (which preserves Lorenz calibration), where $c(k)$ is an arbitrary scalar function, the 3D current can also be replaced with its orthogonal components $\mathbf{j}_\perp$ in formula (\ref{1j}) for the total spin momentum.

It is clear from (\ref{2c1}) that the total number of photons is always non-negative and is explicitly relativistically invariant. It obviously does not depend on time being a constant characteristic of a particular free field configuration -- a property it shares with the total spin momentum (\ref{1j}).

Let us now introduce the following second order tensor
\begin{equation}
\label{2d}
\mathcal{C}^{im}=-\frac{1}{\pi^2}\int\theta(k^0)\delta(k^2)j^i(k){j^m}^*(k)\,d^4k.
\end{equation}
From formulas (\ref{1i}) and (\ref{2c}) it is clear that the total spin momentum and total number of photons can be expressed through $\mathcal{C}^{im}$, respectively, as 
\begin{eqnarray}
S^{im}=\mathrm{Im}\mathcal{C}^{im},\label{2e1}\\
N=\frac{1}{2\hbar}{\mathcal{C}^\nu}_\nu.\label{2e2}
\end{eqnarray}
In other words the total spin momentum and total number of photons can be calculated as the imaginary part and trace of one second order tensor bi-linear in currents. This demonstrates that there is an intrinsic connection between the spin momentum and the number of photons in the classical electromagnetic field whereas the former can be interpreted as the sum of spins of photons present in the electromagnetic field.

To obtain the formulas for the total number of photons and total spin momentum in coordinate representation one should take into account that the following expression
\begin{eqnarray}
\label{2e3}
\int\theta(k^0)&\delta(k^2)e^{-ik_mx^m}\,d^4k=\nonumber\\
&-\frac{2\pi}{x^2}+2i\pi^2\mathrm{sign}(x^0)\delta(x^2)=\nonumber\\
&i(2\pi)^3 D^+(x),
\end{eqnarray} 
where function $D^+(x)$ is known in the quantum field theory as the Wightmen function or positive frequency part of Pauli--Jordan function \cite[p. 219]{rebenko2012theory}. It satisfies the wave equation $\Box D^+(x)=0$. Then the total number of photons and the total spin momentum can be written as 
\begin{eqnarray}
N=&\frac{1}{\pi\hbar}\int\int\frac{j^l(x)j_l(x')}{(x-x')^2}\,d^4x\,d^4x',\label{2e4}\\
S^{im}=&-2\int\int\left[j^i(x)j^m(x')-j^m(x)j^i(x')\right]\times\nonumber\\
&\theta(x^0-x'^0)\delta((x-x')^2)\,d^4x\,d^4x'.\label{2e5}
\end{eqnarray}
In other words the total number of photons and the total spin momentum are auto-correlators of the 4D currents.

In \cite{FeshchenkoVinogradov2018} formula (\ref{2c}) was shown to be equivalent to the Zeldovich formula using coordinate representations  (\ref{2e4})--(\ref{2e5}). The same result can be obtained using momentum representations (\ref{2e4}) and (\ref{2e5}) by introducing 3D Fourier transform of the transverse components of vector potential as 
\begin{equation}
\mathbf{A}_\perp(\mathbf{k})=\frac{2^{3/2}\pi}{|\mathbf{k}|}\mathbf{j}_{\perp}(|\mathbf{k}|,\mathbf{k}).
\label{2e6}
\end{equation}
Using (\ref{2e6}) formula (\ref{2c1}) can be turned into the following expression
\begin{equation}
\label{2e7}
N=\frac{1}{32\pi^4\hbar}\int\frac{|\mathbf{k}|^2|\mathbf{A}_{\perp}(\mathbf{k})|^2}{|\mathbf{k}|}\,d^3k.
\end{equation}
Taking into account that, when $k^0=|\mathbf{k}|$ and Lorentz condition $k^0A^0=\mathbf{kA}$ is satisfied, 3D Fourier transforms of the electrical and magnetic fields can be expressed through $\mathbf{A}_{\perp}$ as 
\begin{eqnarray}
\mathbf{E}(\mathbf{k})=i|\mathbf{k}|\mathbf{A}_{\perp}(\mathbf{k}),\label{2e8}\\
\mathbf{H}(\mathbf{k})=i(\mathbf{k}\times\mathbf{A}_{\perp}(\mathbf{k}))\label{2e9}
\end{eqnarray}
and formula (\ref{2e7}) becomes
\begin{equation}
\label{2e10}
N=\frac{1}{64\pi^4\hbar}\int\frac{|\mathbf{E}(\mathbf{k})|^2+|\mathbf{H}(\mathbf{k})|^2}{|\mathbf{k}|}\,d^3k,
\end{equation}
which is the same Zeldovich formula found at the beginning of \cite{Zeldovich1966}.

\section{Photon number and spin momentum operators}
To make transition to the quantum electrodynamics let us introduce the following two functions of the 3D wave vector $\mathbf{k}$
\begin{eqnarray}
a^-_s(\mathbf{k})=\frac{1}{2\pi\sqrt{\mathbf{k}\hbar}}j_{\perp,s}(|\mathbf{k}|,\mathbf{k}),\label{2f1}\\
a^+_s(\mathbf{k})=\frac{1}{2\pi\sqrt{\mathbf{k}\hbar}}j^*_{\perp,s}(|\mathbf{k}|,\mathbf{k}),\label{2f2}
\end{eqnarray}
where index $s=1,2$ refers to the two non-zero components of $\mathbf{j}_\perp$ orthogonal to $\mathbf{k}$. Now the total number of photons (\ref{2c1}) and the total spin momentum (\ref{1j}) can be expressed as 
\begin{eqnarray}
N=\int\left[a^+_1(\mathbf{k})a^-_1(\mathbf{k})+a^+_2(\mathbf{k})a^-_2(\mathbf{k})\right]\,d^3k,\label{2g1}\\
\mathbf{S}=i\hbar\int\mathbf{n}\left[a^+_2(\mathbf{k})a^-_1(\mathbf{k})-a^+_1(\mathbf{k})a^-_2(\mathbf{k})\right]\,d^3k,\label{2g2}
\end{eqnarray}
where $\mathbf{n}=\mathbf{k}/|\mathbf{k}|$. Formulas (\ref{2g1})--(\ref{2g2}) show that functions $a^-_s$ and $a^+_s$ should be interpreted as annihilation and creation operators of a photon with momentum $\mathbf{k}$ and linear polarization $s$, which will be signified here by adding a hat to them. As they are operators and photons are vector particles with spin being unity $\hat{a}^-_s$ and $\hat{a}^+_s$ must obey a Bose type commutative relation
\begin{equation}
\label{2h}
\{\hat{a}^-_s(\mathbf{k}),\hat{a}^+_{s'}(\mathbf{k}')\}=\delta_{ss'}\delta(\mathbf{k}-\mathbf{k}').
\end{equation}
Substituting functions $a^-_s$ and $a^+_s$ with operators $\hat{a}^-_s$ and $\hat{a}^+_s$ in (\ref{2g1})--(\ref{2g2}) one can get operators $\hat{N}$ and $\hat{\mathbf{S}}$ for the total number of photons and the total spin momentum of the quantum electromagnetic field. However the latter operator does not correspond to any 3D rotation as was shown in\cite{van1994spin}.

As opposed to (\ref{2g1}) expression (\ref{2g2}) is not in diagonal form. It can be diagonalized by a unitary transformation
\begin{eqnarray}
a^-_1(\mathbf{k})=\frac{1}{\sqrt{2}}(b^-_1(\mathbf{k})+b^-_2(\mathbf{k})),\label{2i1}\\
a^+_2(\mathbf{k})=-\frac{i}{\sqrt{2}}(b^+_1(\mathbf{k})-b^+_2(\mathbf{k})),\label{2i2}\\
a^-_2(\mathbf{k})=\frac{i}{\sqrt{2}}(b^-_1(\mathbf{k})-b^-_2(\mathbf{k})),\label{2i3}\\
a^+_1(\mathbf{k})=\frac{1}{\sqrt{2}}(b^+_1(\mathbf{k})+b^+_2(\mathbf{k})),\label{2i4}
\end{eqnarray}
where $b^-_s$ and $b^+_s$ should be again interpreted as annihilation and creation operators of a photon with momentum $\mathbf{k}$ and right ($s=1$) or left ($s=2$) circular polarization and the helicity being equal $1$ or $-1$, respectively. They must obey the same commutative relation (\ref{2h}). Operators $\hat{N}$ and $\hat{\mathbf{S}}$ can finally be expressed through these operators as 
\begin{eqnarray}
\hat{N}=\int\left[\hat{b}^+_1(\mathbf{k})\hat{b}^-_1(\mathbf{k})+\hat{b}^+_2(\mathbf{k})\hat{b}^-_2(\mathbf{k})\right]\,d^3k,\label{2j1}\\
\hat{\mathbf{S}}=\hbar\int\mathbf{n}\left[\hat{b}^+_1(\mathbf{k})\hat{b}^-_1(\mathbf{k})-\hat{b}^+_2(\mathbf{k})\hat{b}^-_2(\mathbf{k})\right]\,d^3k,\label{2j2}
\end{eqnarray}
both being diagonal. 

From (\ref{2g1})--(\ref{2g2}) and (\ref{2j1})--(\ref{2j2}) it is clear that there exists a direct correspondents between the classical expressions for the total number of photons and the total spin momentum and the respective quantum operators. 

\section{Discussion and conclusion}
In this paper we derived a classical expression for the total number of photons in the free electromagnetic field in a systematic way starting from the general solution of the Maxwell field equations. The obtained expression is relativistically invariant by construction and as we have also proved is equivalent to the Zeldovich formula. We also derived the relativistically invariant classical expression for the total spin momentum of the electromagnetic field. We showed that both quantities can be obtained from the same second order tensor bi-linear in current components thus demonstrating that there exists an intrinsic connection between the total photon number and total spin momentum of the field where the latter can be represented as a sum of spins of all photons present in the free field. 

Finally we modified the obtained expressions for the number of photons and spin momentum by introducing two functions of the 3D wave vector, which can be considered as annihilation and creation operators of the photons in the quantum electromagnetic field. These quantized expressions coincide with the known quantum operators for the total number of photons and the total spin momentum of the electromagnetic field demonstrating that there exists a direct correspondence between the quantum and classical definitions of these two quantities.

In conclusion. The concepts of quantum and spin are generally absent in classical physics. However, the classical electromagnetic field has invariants that correspond to the operators of the number of photons and the spin angular momentum known in quantum electrodynamics. In the present paper we conclusively demonstrated that these classical invariants: the total number of photons (Zeldovich formula) and the total spin momentum of the electromagnetic field are limiting cases of the respective quantum operators thus dispelling doubts expressed in a number of previously published papers \cite{avron1999number,mcdonaldlorentz} about the validity of Zeldovich formula. 

\section*{Acknowledgments}
The authors are grateful to Dr. M.A. Vasiliev, Dr. A.E. Shabad and Dr. I.A. Artyukov for fruitful discussions. The work was supported by the Research Programme of the Presidium of the Russian Academy of Sciences \textit{Actual problems of photonics, probing inhomogeneous mediums and materials} PP RAS No 7 as well as by the basic funding within the framework No 0023-0002-2018.

\section*{References}
\bibliography{NunberOfPhotonsEurPhysJ_Feshchenko} 
\bibliographystyle{unsrt} 

\end{document}